\documentstyle[preprint,aps,epsf]{revtex}                  
\begin{document}
\pagestyle{empty}                                      
\preprint{
\font\fortssbx=cmssbx10 scaled \magstep2
\hbox to \hsize{
\hfill$\raise .5cm\vtop{              
                \hbox{NTUTH-97-01}\hbox{NCTU-TH-97-01}\hbox{MSUHEP-70122}}$}
}
\draft
\vfill
\title{Probing Flavor Changing Neutral Higgs Couplings at LHC
}

\vfill
\author{Wei-Shu Hou$^a$, Guey-Lin Lin$^b$, Chien-Yi Ma$^{c}$ and C.--P. Yuan$^d$}
\address{
$^a$\rm Department of Physics, National Taiwan University,
Taipei, Taiwan, R.O.C.
}
%
%
\address{
$^b$\rm Institute of Physics, National Chiao Tung University,
Hsinchu, Taiwan, R.O.C.
}
\address{ $^c$\rm Department of  Electrophysics, National Chiao-Tung University, Hsinchu, 
Taiwan
, R.O.C.}
\address{$^d$\rm Department of Physics and Astronomy, Michigan State University, 
East Lansing, \\
MI 48824, U.S.A.}

%
%
\vfill
\maketitle
\begin{abstract}
Uniquely heavy with mass at the weak scale, the top quark may have
large flavor changing couplings to Higgs bosons that are as yet unexplored.
We show that such couplings
can be {\it directly} probed at the LHC through the parton subprocess
$c(\bar c)g \to t (\bar t)A^0$, where the pseudoscalar $A^0$
subsequently decays into $t \bar c$ or ${\bar t} c$,
giving rise to the intriguing final state of like sign top quark pairs.
After demanding $\ell^\pm\ell^\pm$, missing energy and two $b$-jets,
the major background turns out to be $q\bar q' \to Wt \bar t$,
which can be partially suppressed by jet counting.
The signal can then manifest itself in the asymmetry of numbers of
$\ell^+\ell^+$ and $\ell^-\ell^-$ events.
To further improve the signal over background,
efficient $t$ vs. $\bar t$ tagging methods should be developed.

\end{abstract}
%
%
%
%
\pagestyle{plain}


Despite spectacular agreement with experiment,  the Standard Model  (SM)
offers limited insight into its own structure. 
In particular,  it does not explain but simply parametrizes the hierarchical patterns
seen in both  the fermion masses and the CKM mixing  matrix.
Nor does it reveal any details of the Higgs sector
which is responsible for the electroweak symmetry breaking.
Before a theory to account for these issues can be established,  
it is essential to gather experimental hints by exploring
the properties of  the top quark and the Higgs boson(s). 
Since $m_t$ is of  the same order as the  electroweak symmetry breaking scale,  
flavor dynamics involving the top quark and 
the  electroweak symmetry breaking mechanism
might be closely related to each other.  
In this note we examine the case where the top quark possesses
large flavor changing couplings to neutral Higgs bosons \cite{Hou,HW}.

To focus our discussions, we adopt the scenario that 
electroweak symmetry breaking is driven by a scalar sector.  
Given stringent limits on flavor changing neutral current (FCNC) processes,
the structure of this scalar sector beyond its simplest form in SM is very restricted.     
In multi-Higgs doublet models, it is customary 
to impose some discrete symmetries \cite{GW} to ensure the absence of 
flavor changing neutral Higgs couplings (FCNH) at the tree level. 
However, inspired by the quark mass and mixing hierarchy pattern,
\begin{equation}
   \begin{array}{ccccccc}
    m_1 & \ll & m_2 & \ll & m_3, && \\
    \vert V_{ub}\vert^2 & \ll & \vert V_{cb} \vert^2 &
                      \ll & \vert V_{us}\vert^2 & \ll &1,
   \end{array}
\end{equation}
Cheng and Sher \cite{CS} suggested that low energy FCNC 
can be naturally suppressed without invoking discrete symmetries.
For example, in the general two-Higgs doublet model (2HDM),
quark mass matrices consists of  two parts,
{\boldmath $m$} $ = $ {\boldmath $m$}$^{\left(1\right)} 
                               + ${\boldmath $m$}$^{\left(2\right)}$,
where
{\boldmath $m$}$^{\left(1\right)}$ and {\boldmath $m$}$^{\left(2\right)}$ 
are each induced by vacuum expectation values (v.e.v.): 
$v_1=\langle\Phi_1^0\rangle$ and $v_2=\langle\Phi_2^0\rangle$. 
To sustain   (1), unless fine-tuned cancellations occur,
the off-diagonal elements of {\boldmath $m$}$^{\left(1\right)}$
and {\boldmath $m$}$^{\left(2\right)}$, just like {\boldmath $m$} itself,
should trickle off as one moves off-diagonal. 
Hence, the FCNH coupling matrices {\boldmath $\xi$}$^{\left(k\right)}$,
obtained from $\sqrt{2}${\boldmath $m$}$^{\left(k\right)}/v_k$ 
by rotating to the mass eigenbasis, cannot be arbitrary.  
Based upon this observation, Cheng and Sher proposed \cite{CS} 
the ansatz      
$\xi^{\left(k\right)}_{ij} \sim {\sqrt{m_i m_j}/v_k}$.
Thus, FCNH couplings involving lower generation fermions 
are naturally suppressed, without pushing
FCNH  Higgs boson masses to way beyond the v.e.v. scale. 
However, since $\sqrt{2} m_t\cong v\equiv \sqrt{v_1^2 + v_2^2}$,  
the flavor  changing coupling  $\xi^{\left(k\right)}_{tc}$ could be quite sizable, 
and could hence lead to interesting consequences
such as $t\to c+ S^0$ \cite{Hou,HW} or $S^0\to t\bar c,\ \bar tc$ \cite{Hou},
where $S^0$ is some neutral Higgs boson. 
Since top decay seems to proceed predominantly via $t\to bW^+$,
we shall be interested in the case where neutral Higgs bosons
are heavier than the top quark.

Before we go on to discuss how to probe $\xi^{\left(k\right )}_{tc}$,
let us rotate $\Phi_1$ and $\Phi_2$ \cite{LS} 
such that  $\left\langle \phi_2^0 \right\rangle =0$ and
$\left\langle \phi_1^0 \right\rangle =v/ \sqrt{2}$. 
This eliminates $\xi^{\left (1\right)}$ and transforms  $\xi^{\left (2\right)}$ into
\begin{equation}
\xi_{ij} =f_{ij} {\sqrt{m_i m_j}/v},
\end{equation}
where $f_{ij}$'s are constants of order unity.
In this new basis, 
the pseudoscalar $A^0 \equiv \sqrt{2}\, \mbox{Im}\, \phi_2^0$
and charged scalar $H^\pm \equiv \phi_2^\pm$ are physical Higgs bosons.
The CP even neutral scalars 
$\sqrt{2}\, \mbox{Re}\,\phi_1^0$ and $\sqrt{2}\, \mbox{Re}\,\phi_2^0$
mix through the Higgs potential into the physical states $H^0$ and $h^0$.
In the limit that the mixing angle $\sin\alpha \to 0$,
$H^0 \leadsto \sqrt{2}\, \mbox{Re}\, \phi_1^0$
becomes the ``standard" Higgs boson with
diagonal couplings,
while
$h^0 \leadsto \sqrt{2}\, \mbox{Re}\, \phi_2^0$
has Yukawa couplings as in   (2),
but decouples from vector
or $H^+H^-$ boson pairs, just like $A^0$.

For $m_{A^0} < m_{H^+} + M_W$ and $m_{h^0/H^0} + M_{Z^0}$
(easily realized if all Higgs bosons have mass $\sim v$),
and assuming that CP is a good approximate symmetry,
the $A^0$ is unique in that it {\it decays only into fermionic final states}.
Between the $t\bar c$ and $t\bar t$ thresholds,  
\begin{equation}
200\ \mbox{GeV} < m_{A^0} < 2m_t \simeq 350\ \mbox{GeV},
\end{equation}
with the coupling of   (2), 
{\it $A^0 \to  t\bar c,\ \bar tc$ would dominate over
the usual $b\bar b$ mode} and become dominant \cite{Hou}.
We shall therefore concentrate on $A^0$ 
in this mass range as an FCNH probe.

The signatures of sizable $\xi_{tc}$ coupling 
at $\ell^+\ell^-$ colliders 
have been discussed recently \cite{ARS,HL}. 
In particular,  it was suggested \cite{HL}
that at  the 500 GeV Next Linear Collider (NLC),
the process $e^+e^- \to Z^*\to h^0 A^0 \to t\bar c\; t \bar c$
or $\bar tc\;  \bar tc$
could lead to an intriguing final state with
{\it like sign top quark pairs},
the traditional hallmark of neutral meson mixings
(note that $T_q = t\bar q$ mesons do not even form).  
With both $m_{h^0}$ and $m_{A^0}$ in range of (3), and if  $\sin\alpha\to 0$, 
the $tt\bar c\bar c$ ($\bar t\bar tcc$) final state is most  favorable since both
$h^0$ and $A^0$ dominantly decays into $t\bar{ c }\ (\bar{t} c)$.  
However, with an integrated luminosity of 50 fb$^{-1}$,  
one expects no more than a score of 
like sign dilepton events at NLC per year,
which improves if $\sqrt{s}$ can be raised.
For generic $\sin\alpha$ values where 
$h^0\to WW,\ ZZ$ decays are dominant, 
the like sign top final state is no longer favorable,
but  like sign dilepton events are still expected from 
the  $W^+W^-t\bar{c}(\bar{t}c)$ final state. 
Note that the strength of $\xi_{tc}$ is only {\it indirectly} probed
through the {\it decays} of $h^0$ and $A^0$ here. 
To probe $\xi_{tc}$ directly and with a larger event rate, we turn to hadronic colliders.
 
It is instructive to compare the event rates of $h^0A^0$ production 
at linear and  hadronic colliders.
At Tevatron, one expects no gain for $q\bar{q} \to h^0 A^0$ 
since production cross section is similar,
but the luminosity at a few fb$^{-1}$ per year  with the Main Injector 
is smaller than at the NLC. 
The situation is only slightly improved at the LHC. The cross section 
of $h^0A^0$ production is about 7 fb for $m_{A^0}=m_{h^0}=250$ GeV.  
With an integrated luminosity of 100 fb$^{-1}$, 
there are 700 pairs of $h^0A^0$ produced each year,  
which is not much better than $\sim 300$  $h^0A^0$ pairs \cite{HL}
produced at an NLC with energy extended to 600 GeV (assuming 50 fb$^{-1}$),
but facing more backgrounds.   

Hadronic colliders, however, offer the opportunity to involve 
the strong interaction in the {\it production process}, 
which can be used to {\it directly probe  $\xi_{tc}$}.  
Surveying  $q\bar{q},\ qg$ and $gg$ processes,
to have $\xi_{tc}$ appearing in one of the interaction vertices,
in general one requires $2\to 3$ scattering, such as
$q\bar{q}\to g^*\to t\bar{c}(\bar{t} c)A^0$.  
However, the cross sections turn out to be very small,
and it would be advantageous if $2\to 2$ scattering is possible.
We find that the $c(\bar {c})g\to t(\bar{t})A^0$ process
is rather promising in this regard as a direct probe to  $\xi_{tc}$. 
Although the cross section at Tevatron remains small, 
the situation changes drastically at LHC.

At Tevatron energies, to produce an $A^0$ of 250 GeV 
in association with a top quark, the colliding partons must carry
large momentum fractions, hence both charm and gluon 
distribution functions are very suppressed, 
resulting in a very small $tA^0$ production cross section. 
>From  Fig. 1 and using CTEQ3L \cite{CTEQ3} parton distribution functions, 
the cross section at the Tevatron is only about $10^{-2} f^2 $ fb for $m_{A^0}=250$ GeV,
where $f=f_{tc}$ is the constant appearing in   (2). 
Though the cross section is very small, it is proportional to $f^2$
hence a direct probe to FCNH coupling $\xi_{tc}$.  
At the LHC with $\sqrt{s}=14$ TeV, 
the colliding parton momentum fractions could be much smaller 
so that both charm and gluon distribution functions contribute significantly. 
Repeating the calculation for LHC, with $m_{A^0}$= 250 GeV
we obtain a cross section of $37f^2$ fb which is 
3000 times larger than that at the Tevatron,
one order of magnitude larger than 
$h^0A^0$ associated production, and grows as $f^2$.
We show in Fig. 2 the dependence of  $\sigma (pp\to t(\bar{t})A^0+X)$
on $m_{A^0}$ with $f$ taken to be unity.  

With $A^0\to VV$ forbidden by CP invariance,
$A^0$ decays predominantly into $t\bar{c}$ or $\bar{t}c$ 
in the mass range given by   (3).
For example, for $m_{A^0}=250$ GeV, 
$90\%$ (fraction increases with $m_{A^0}$) of  $A^0$ 
decays into the above final states \cite{HL}, half of which 
pair up with the associated top to make a like sign top pair event.
The signature of such events are like sign dileptons,
accompanied by two $b$-jets, large missing energy,
plus one additional jet,
\begin{equation}
cg\to tA^0\to \ell_1^+\ell_2^+\nu\nu + bb + \bar{c},
\end{equation}
and similarly for
$\bar{c}g\to \bar{t}A^0\to \ell_1^-\ell_2^- \bar{\nu}\bar{\nu} + \bar{b}\bar{b} + c$.
With an integrated luminosity of 100 fb$^{-1}$ 
and 50$\%$ double $b$-tagging efficiency,
we expect for both $\ell^+\ell^+$ and $\ell^-\ell^-$ modes
\begin{equation}
 37f^2\times {90\%\over 2}\times {4\over 81} \times 50\% \times 100=40f^2
\end{equation} 
events per year for $m_{A^0}=250$ GeV.
The event rate for other values of $m_{A^0}$ can be read off from Fig. 2,
together with Fig. 1 of  Ref. \cite{HL}, 
where the $m_{A^0}$ dependence of  BR($A^0\to t\bar{c}+\bar{t} c$) is plotted. 
Over the mass range of   (3), the event rate does not change significantly
since $m_{A^0}$ affects the production cross section 
and BR($A^0\to t\bar{c}+\bar{t} c$) in compensating ways.

One might think that the same final state may also be reached by 
single top production followed by $A^0$ bremsstrahlung, as shown in Fig. 3. 
This is analogous to the production of a Higgs boson 
associated with a single top \cite{WG}. 
In the current  context, we have
\begin{equation}
qb\to q' t A^0.
\end{equation} 
Since the simpler parent process, 
the so-called single-top production $qb\to q't$,
has a cross section around 100 pb \cite{CY},
the process (6) would appear to have a  large cross section.  
Adding an $A^0$ to the final state tends to 
reduce the cross section by 3 orders of magnitude, 
but a cross section for $qb\to q' t A^0$ around 100 fb is still quite large.

To ascertain this, we divide the total cross section into three parts,
$\sigma =\sigma_t + \sigma_{H^+} +\sigma_{tH^+}$,
where the first two terms are from each diagram alone,
and the third is their interference.
With CTEQ3L \cite{CTEQ3} parton distribution functions,
we find $\sigma_t$, $\sigma_{H^+}$,  and $\sigma_{tH^+}$ 
to be 21.7, 24.4 and  $-$43.6 fb, respectively, 
for $m_{H^+} = m_{A^0} = $ 250 GeV. 
The interference term almost cancels the diagonal terms completely
and renders a total cross section of only 2.5 fb, 
which is much smaller than $cg \to tA^0$! 
The result is found to be not very sensitive to $m_{H^+}$ for $m_{H^+} > m_{A^0}$,
and was double checked with helicity methods \cite{CY}.
Such cancellation is pretty much a consequence of unitarity,
and
is rather analogous to the SM $q b \to q' tH^0$ case \cite{WG}.  

We have now singled out $cg\to tA^0$ as the most promising  
mode to probe the FCNH coupling $\xi_{tc}$.  
This is a {\it direct probe of $\xi_{tc}$},
since $f$ can be determined from the cross section of $cg\to tA^0$. 
The method depends only on the mass of $A^0$ 
and the decay branching ratio for $A^0 \rightarrow t {\bar c} ({\bar t} c)$ to give
like-sign top quark pair events.
Furthermore, because $A^0$ is CP-odd, the above decay branching ratio 
is almost model-independent for the values of $m_{A^0}$ condsidered.
What remains to be checked are the backgrounds. 
We focus on like sign $W$ pair production,
which would also give rise to  like sign dilepton events.

Vector boson pair production has been studied extensively \cite{BB}
for the purpose of  probing the electroweak symmetry breaking mechanisms. 
By requiring two $b$-jets and like sign dileptons in the final state, 
one can discard all of these,
{\it except} 
$q\bar{q'} \to W^+(W^-)t\bar{t}$ \cite{KUN} (see Fig. 4).  
The production and decay chain
\begin{equation}
u\bar{d}\to W^+ t\bar{t}\to W^+W^+W^-b\bar{b}
\to \ell_1^+\ell_2^+\nu\nu + b\bar{b} + j_1j_2,
\end{equation}
leads to like sign dileptons
as well as a pair of $b$- and $\bar b$-jets
(and likewise for 
$d\bar{u}\to W^- t\bar{t}\to W^-W^-W^+b\bar{b}\to
\ell_1^-\ell_2^-\bar{\nu}\bar{\nu} + b\bar{b} + j_1j_2$).
Unlike the signal process of (4), there are {\it two jets} $j_1$ and $j_2$,
which should have pair mass $m_{jj}$ around $M_W$.
Convoluting with parton distribution functions, 
we find $\sigma ( pp \to W^+ t\bar{t} +X)=210$ fb 
while $\sigma (pp \to W^- t\bar{t} +X)=100$ fb, 
which agrees with the results of Barger et al.,
ref. \cite{KUN}.
The factor of two
comes from the dominance of valence contributions, 
i.e. $u(x)=2d(x)$ and $\bar{d}(x)= \bar{u}(x)$ for $pp$ initial state.
   
Assuming100 fb$^{-1}$ per year at the LHC, 
the annual event number for process (7) is
\begin{equation}
210\times {2\over 3}\times {4\over 81}\times 50\% \times 100=350,
\end{equation}
and half this rate for $\ell^-\ell^- + X$ events.
The factor  of ${2/3}$ is the $W\to jj$ branching ratio.  
The background of (8) appears to dominate over the signal of (5) 
both in $\ell^+\ell^+$ and $\ell^-\ell^-$ modes, 
though it is less severe in the latter case.  
Adding to the problem, we find that the $W$ boson associated with the $t\bar t$ pair  
also turns out to be produced in the central region,
hence a Monte Carlo study is needed 
to separate signal from background.
While details of such a study will be presented elsewhere, 
let us provide a qualitative argument on this matter. 
The simplest  way is clearly jet counting. 
Two $b$-jets are already tagged, 
but it may be too costly to determine $b$ vs. $\bar b$.
The signal has one additional jet  
while the background has two, with $m_{jj} \simeq M_W$. 
If the two background jets are both in the central region 
($\vert\eta\vert < 3$, where $\eta$ is pseudorapidity) 
and can be distinguished, 
the event can be excluded by na\"\i ve jet counting. 
If the two jets merge into one large jet $J$, 
the event can still be effectively removed
by cutting on large $m_J$ around $M_W$. 
Only if either $j_1$ or $j_2$ falls outside of the detection region
or coalesce accidentally with one of the $b$-jets
will the event become an irreducible background. 
This kinematics is however unlikely because the $t\bar t$ syetem,
which gives rise to $W\to j_1 j_2$, is centrally produced
as discussed before.
A conservative estimate is that by  jet counting  alone, 
one should be able to reduce the background by at least 50$\%$ \cite{GPY}.

With simple jet counting, one has less than 90   
background $\ell^-\ell^-$ events, 
with an excess of  $\sim 40f^2$ coming from signal events. 
For $f\sim \sqrt{2}$, signal and background event rates
would be comparable.
In other words, if  the FCNH coupling 
$\xi_{tc} = f {\sqrt{m_c m_t}/v}$ indeed exists, 
considerable excess shall be observed in $\ell^-\ell^-$ events.
For a slightly larger $f$, say $f\sim 2$, the signal is also 
comparable to the background in the  $\ell^+\ell^+$ mode. 
Defining $N(\ell^\pm\ell^\pm)$ as the number of  $\ell^\pm\ell^\pm$ events, 
the signal can then manifest itself in the asymmetry parameter
\begin{equation}
 A={N(\ell^+\ell^+)-N( \ell^-\ell^-)\over N(\ell^+\ell^+)+N( \ell^-\ell^-) }.
\end{equation} 
The background alone gives $A={1\over 3}$, 
while the signal events lower $A$ to ${1\over 7}$ for $f\sim 2$.

Though interesting in itself, it should be noted that
$q\bar q^\prime \to Wt\bar t$ enters as background only because 
it is difficult to {\it tag the top flavor} at present. 
To demand same-flavor $b$ tag might be too costly in terms of efficiency.
The signal to background ratio would be greatly improved 
once $t$ vs. $\bar{t}$ can be easily distinguished  experimentally.
The development of such techniques should  be pursued with priority
at the LHC and for the longer term future,
since flavor and CP violation are closely linked.
Eventually we would  like to study top flavor violation 
as a probe of CP violation in $t\bar t$ production and decay.

We stress that $cg\to tA^0$ can be viewed as 
a model independent probe to FCNH couplings. 
To produce like sign top events, only a sizable $A^0tc$ coupling $\xi_{tc}$ 
and a fairly large branching ratio for  $A^0 \to t\bar{c}+\bar{t}c$ are essential. 
If  the first condition is satisfied, it is very probable that the second holds as well.  
One simply needs to argue that  the branching ratio for two boson decays,
$A^0\to VV$, is suppressed. 
This should indeed be the case
since CP appears to be a good approximate symmetry, 
and $A^0VV$ coupling 
can only be generated by loop corrections.
Moving into the model independent realm,
we note in passing that $ug\to tA^0$ could be much more
prominent than discussed here, {\it if} $\xi_{ut}$ is comparable to
$\xi_{ct} \sim {\sqrt{2m_c m_t}/v} \sim 0.1$.
In this case the large valence distribution function
leads to a $tA^0$ cross section of order 500 fb for $m_{A^0} = 250$ GeV,
but with little gain for $\bar tA^0$.
As a final remark, we note that $gg\to A^0 \to t\bar c$ cross section
is considerably larger than those discussed here,
but would be plagued by the much larger single top and $t\bar t$
cross sections at the LHC. 

In summary, we have shown that FCNH couplings can be directly probed at the LHC 
via like sign top quark pair production through the $cg\to tA^0\to tt\bar q$ process. 
Possible backgrounds are identified and calculated.
 To better distinguish the signal from backgrounds, 
the results of a detailed Monte Carlo study will be reported in a future publication. 
Apart from this, it is also essential  to calculate 
the background $q\bar{q'}\to W^\pm t\bar{t}$ more accurately, 
in particular to the NLO accuracy where dependence on 
the renormalization (factorization) scale could be significantly reduced.
More efficient top flavor tagging ($t$ vs. $\bar t$!) methods are desirable 
to convincingly observe the possible production of like-sign top quark pairs.

\acknowledgments
We thank Darwin Chang, Jyh-Liong Lim and Gong-Ping Yeh  for useful discussions.
The works of WSH, GLL and CYM are supported in part by 
National Science Council of R.O.C. under grant numbers 
NSC 86-2112-M-009-026 and NSC 86-2112-M-009-012.
CPY is supported in part by NSF grant No. PHY-9507683.  
%


\begin{figure}
\caption{Subprocess $cg\to tA^0$.}
\label{fig1}
\end{figure}

\begin{figure}
\caption{Cross section for $pp\to tA^0 + X$ at LHC via subprocess of Fig. 1.}
\label{fig2}
\end{figure}

\begin{figure}
\caption{Subprocess $qb\to q^\prime t A^0$.
}
\label{fig3}
\end{figure}

\begin{figure}
\caption{Standard model $q\bar q^\prime \to Wt\bar t$ subprocess..
}
\label{fig4}
\end{figure}

\vfill\eject
\epsfxsize=0pt
\centerline{\epsfbox{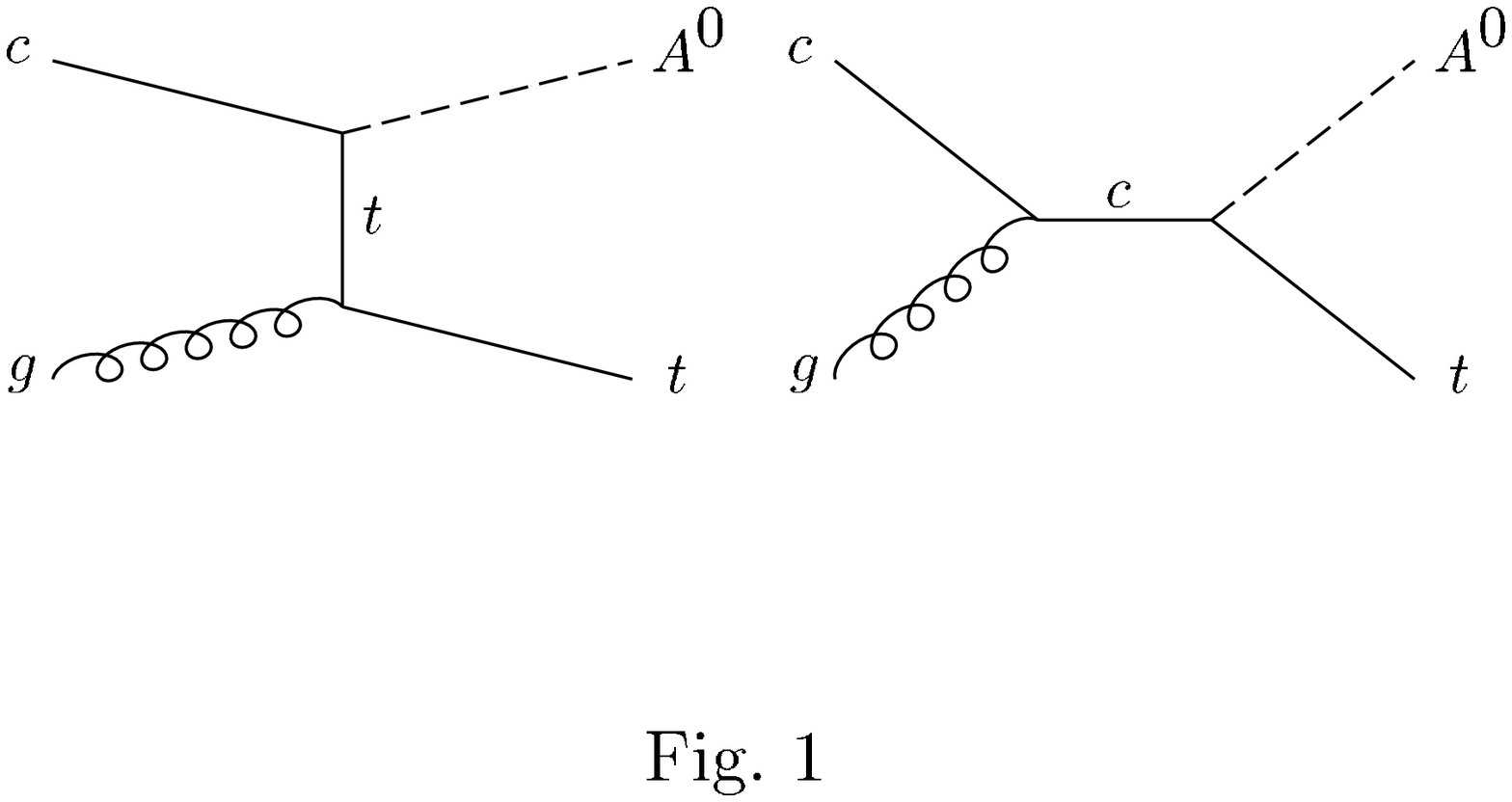}}
\vskip 1cm
\vfill\eject
\epsfxsize=0pt
\centerline{\epsfbox{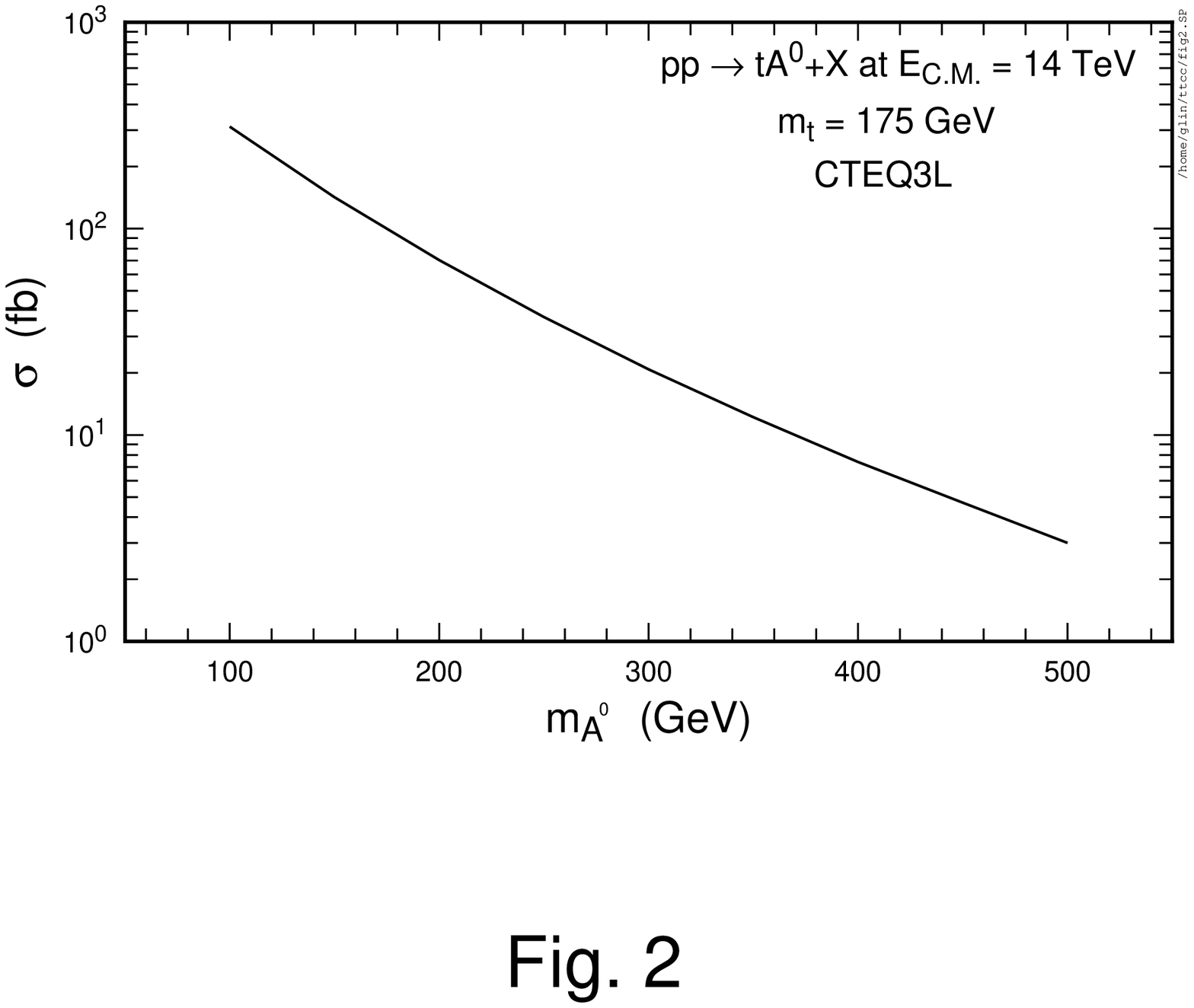}}
\vskip 1cm
\vfill\eject
\epsfxsize=0pt
\centerline{\epsfbox{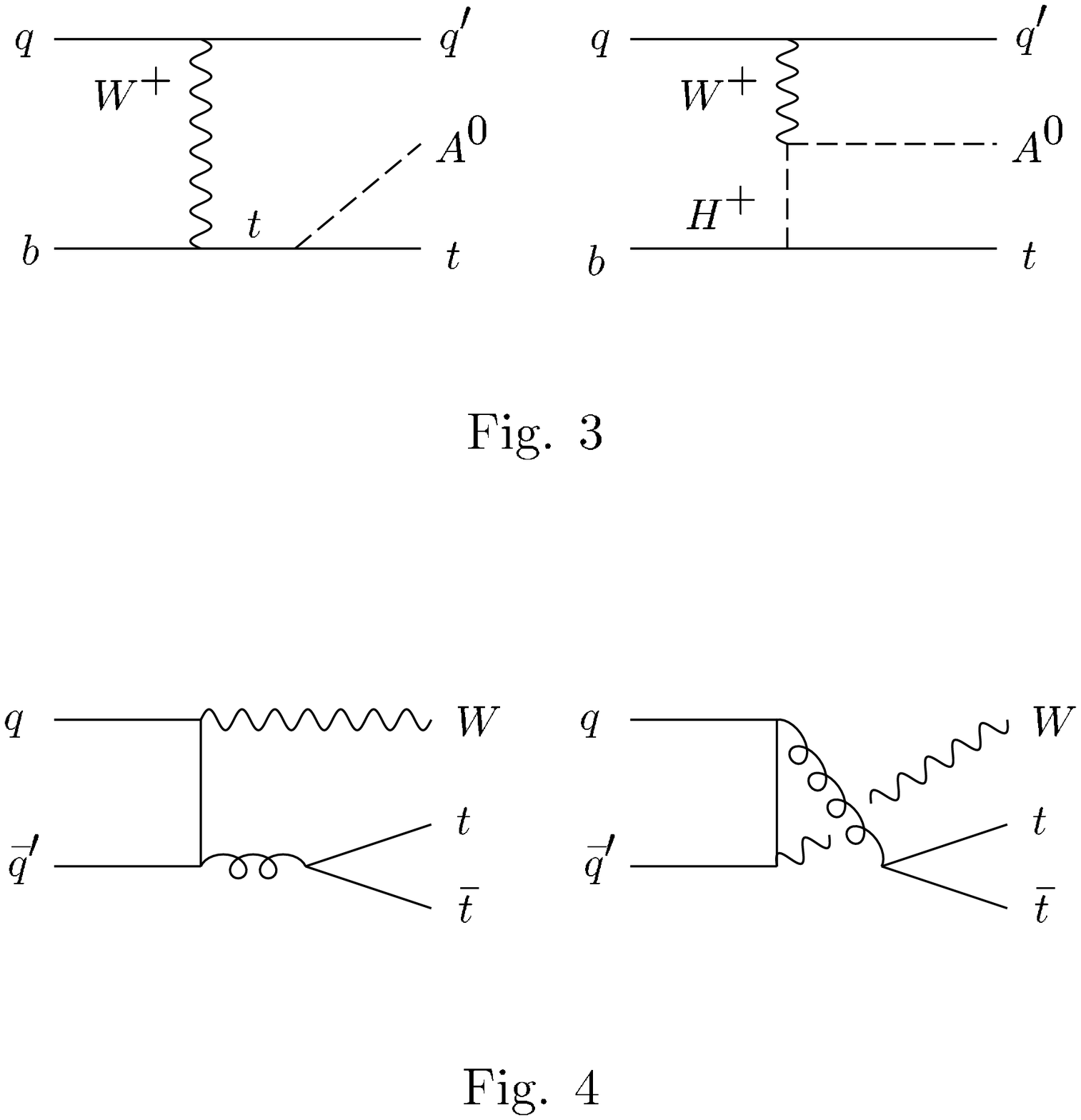}}
\vskip 1cm
\vfill\eject

\end{document}